\documentclass[fleqn,usenatbib]{mnras}
\usepackage[utf8]{inputenc}
\usepackage{mathrsfs}
\usepackage{natbib}
\usepackage{amsmath}
\usepackage{amsfonts}
\usepackage{amssymb}
\usepackage{graphicx}
\usepackage{hyperref}
\usepackage[usenames,dvipsnames]{color}
\usepackage{mathrsfs}

\makeatletter

\usepackage{mathrsfs}

\newcommand{\abs}[1]{\left\vert#1\right\vert}

\newcommand{\arccosh}{\mathrm{arccosh}}

\newcommand{\scri}{\mathscr{I}}

\makeatother

\title[External potential and binaries softening]{Binaries are softer than they seem: Effects of an external potential on the scattering dynamics of binaries}

\author[Y. B. Ginat and H. B. Perets]{Yonadav Barry Ginat,$^{1}$\thanks{E-mail: ginat@campus.technion.ac.il}
and
Hagai B. Perets$^{1}$
\\
$^{1}$Faculty of Physics, Technion -- Israel Institute of Technology,
Haifa, 3200003, Israel
}

\date{Accepted XXX. Received YYY; in original form ZZZ}

\pubyear{2021}
\begin{document}
\label{firstpage}
\pagerange{\pageref{firstpage}--\pageref{lastpage}}
\maketitle

\begin{abstract}
Binary evolution is influenced by dynamical scattering with other stars in dense environments. Heggie's law states that, due to their environments, hard binaries (whose orbital energy surpasses the typical energy of single stars) tend to harden (increase their orbital energy), while soft binaries typically soften. Here, we show that Heggie's law sometimes needs to be revised, by accounting for an external potential, for example, for binaries in nuclear stellar discs or AGN discs, that are affected by the central massive black hole, or binary planetesimals in proto-planetary discs, affected by the host star.
We find that in such environments, where the Hill radius is finite, binary-single scattering can have different outcomes. In particular, a three-body encounter could be cut short due to stars being ejected beyond the Hill radius, thereby ceasing to participate in further close interactions. This leads to a systematic difference in the energy changes brought about by the encounter, and in particular slows binary hardening, and even causes some hard binaries to soften, on average, rather than harden.
We use our previously derived analytical, statistical solution to the bound chaotic three-body problem to quantitatively characterise the revision of the hardening-softening phase transition and evolution of binaries. We also provide an analytical calculation of the mean hardening rate of binaries in any environment (also reproducing the results of detailed N-body simulations). We show that the latter exhibits a non-trivial dependence on the Hill radius induced by the environment.
\end{abstract}




\section{Introduction}
Binaries residing in dense regions such as stellar clusters evolve by interacting with other stars. Such interactions proceed by three-body scattering of a lone star off the binary, which is a chaotic process if the initial binary was `hard' (i.e., its binding energy was much greater than the typical energy of the random motions in the cluster; see, e.g. \citealt{Heggie1975,Hills1975,HeggieHut2003}).
Because these interactions are chaotic, past modelling relied heavily on numerical simulations \citep{Saslawetal1974,Hills1975,HillsFullerton1980,Anosova1986,AnosovaOrlov1986,Hills1989,Hills1992,HeggieHut1993,Hut1993,SigurdssonPhinney1993,Mikkola1994,Samsingetal2014,
LeighWegsman2018,Manwadkaretal2020,Manwadkaretal2021}, but they were also modelled statistically, under certain approximations \citep{Heggie1975,Monaghan1976b,NashMonaghan1978,ValtonenKarttunen2006,Kol2020}.

Recently a probabilistic, closed-form solution of the cross-section for the unbound three-body problem was found by \cite{StoneLeigh2019}, followed by an analogous generalised formula for both the bound and the unbound problem \citep{GinatPerets2020}. When a close three-body interaction terminates, one of the three stars is ejected: in the bound case, it has negative energy, so it has to return. In the mean time, however, the triple is in a hierarchical state; such orbits, therefore, have a pericentre which is of a similar scale as the semi-major axis (sma) of the remnant (inner) binary, but an apocentre much larger than it. Eventually, though, even on a bound orbit, this apocentre can be so large, that the force on the third star from other stars in the cluster (or other external potentials, e.g. Galactic tides in the field, or a massive black hole in galactic nuclei) surpasses the gravitational force induced by the remnant binary, whereupon the star ceases to be bound to it. Thus, the binary-single encounter may terminate even when the energy of the third star is negative, necessitating the use of the statistical solution to the bound three-body problem, which naturally incorporates such a cut-off \citep{GinatPerets2020}.

\cite{GinatPerets2020} found the probability distribution for a final binary energy, following a hard-binary-single encounter, given a total energy $E$ and total angular momentum $J$, to be well-approximated by
\begin{equation}\label{eqn:f_bin E_b}
    f(E_b|E,J) \propto m_b \frac{\theta_{ap}(E_b,R) \scri(E_b,R)}{\abs{E_b}^{3/2}\abs{E - E_b}^{3/2}},
\end{equation}
where $m_b$ is the mass of the remaining binary, $E_b$ is its energy, $\scri$ is a function of $E_b$, that goes like $\abs{E_b}^{-1}$ for large $J$, but like $\abs{E_b}^{-1/2}$ for low $J$, and
\begin{equation}
    \theta_{ap}(E_b) \equiv \begin{cases}
                                                     \arccos\left(1-\frac{R}{a_s}\right) - \sqrt{2\frac{R}{a_s} - \frac{R^2}{a_s^2}}, & \mbox{bound} \\
                                                     \sqrt{2\frac{R}{a_s} + \frac{R^2}{a_s^2}} - \arccosh\left(1+\frac{R}{a_s}\right), & \mbox{unbound}.
                                                   \end{cases},
\end{equation}
where $a_s$ is the semi-major axis of the ejected star relative to the binary, and $R$ is the minimum distance that the third star can have from to the binary's centre of mass, and still form a hierarchical structure.

In this paper, we investigate the effects of introducing a cut-off in more detail, and comment on its influence on the evolution of binaries in various astrophysical environments of interest. In particular, if the cut-off on the maximum distance that the single star can have, while still being bound to the binary, and therefore still be in the middle of a triple interaction, is sufficiently stringent, it is theoretically possible for formally `hard' binaries to soften with time, due to encounters with single stars in their environment, rather than harden, as one would na\"{i}vely assume from Heggie's law. The introduction of a finite cut-off serves, therefore, as a modification of this law; it is small in many cases, but in certain circumstances (e.g. disc environments) it can be quite large. We start with a simple theoretical computation of the effects of such a cut-off, and then move on to explore its implications for various environments, in particular, disc environments, where its effects are larger.


\section{External environmental cut-off}
The introduction of a cut-off can have drastic consequences for the probability that the final binary becomes more energetic (i.e. `harder') or less energetic (`softer'). Heggie's law \citep{Heggie1975} dictates that hard binaries tend to harden, while soft binaries tend to soften. Taking into account the external potential of a cluster might change this: now, the ejected star does not need to have strictly positive energy in order to escape from the triple -- only more than the (already negative) background potential. By the virial theorem, this background potential is about the same order of magnitude as the typical amount of energy that the in-coming perturber has initially, so it is possible, even for hard binaries, to become softer.

\begin{figure*}
    \centering
    \includegraphics[width=0.45\textwidth]{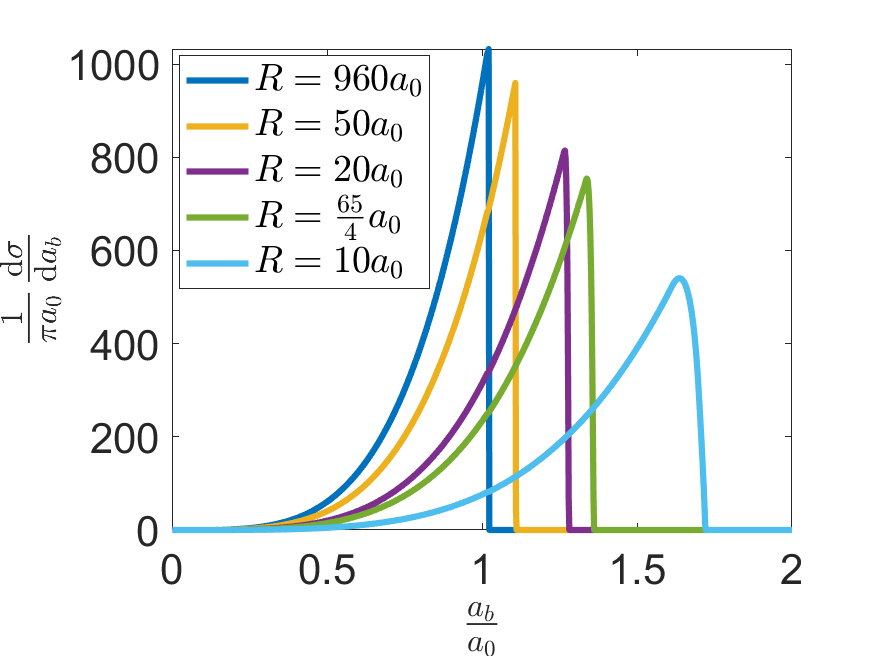}
    \includegraphics[width=0.45\textwidth]{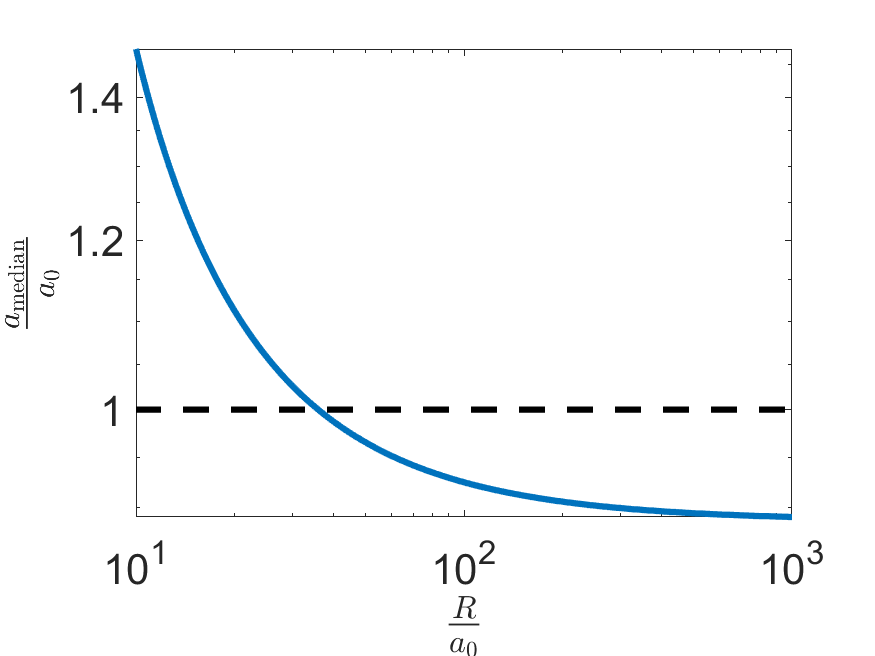}
    \caption{Left: a plot of the differential cross-section for having a final semi-major axis $a_b$ for various choices of $R_{\max}$ (here shortened to $R$). The masses are $m_1 = m_2 = m_3 = 1~M_\odot$, and $a_0 = $ 1 AU, $e_0 = 0$, and the in-coming perturber's initial relative velocity is $v_0 = 5~\textrm{km s}^{-1}$. Right: a plot of the median value of $a_b$ as a function of $R_{\max}$. For small values of $R_{\max}$, $a_{\rm median}$ is larger than $a_0$, so the binary becomes softer in probability; for large $R_{\max}$, $a_{\rm median} < a_0$, and the binary hardens as one would expect.}
    \label{fig:various R}
\end{figure*}

As in \cite{GinatPerets2020}, we model the dependence on the background potential by introducing a cut-off to the maximum distance between the ejected particle on a hierarchical orbit, and the remaining binary, which we shall denote by $R_{\max}$. Clearly, if $R_{\max}$ is too small (relative to the initial semi-major axis $a_0$), the remaining binary will have a greater probability for becoming softer, than for becoming harder. Conversely, if $R_{\max} = \infty$, the opposite holds (see, e.g. \citealt{Heggie1975,HeggieHut2003,Binney}). Let us calculate the critical value of $R_{\max}$, where the probabilities are equal to each other; for simplicity, we assume that all stars are identical, with mass $m$. We plot the differential cross-section for having a final semi-major axis $a_b$ for various choices of $R$ in figure \ref{fig:various R}. In this paper, we always take $m_1$ and $m_2$ to be the masses of the initial binary members, and $m_3$ to be the perturber's mass.

The probability density function for the inner binary to have a semi-major axis $a_b$ is well-approximated by \citep{GinatPerets2020}
\begin{equation}
  f(a_b) \propto \begin{cases}
                   \scri(a_b) a_b, & \mbox{if } a_b < a_{\rm co} \\
                   0, & \mbox{otherwise}.
                 \end{cases}
\end{equation}
where $a_{\rm co}$ is a cut-off, related to $R_{\max}$, as will be explained below, and $\scri$ is angular-momentum related integral, which is well-approximated\footnote{Here we wish to keep the derivations as simple as possible, but see \cite{GinatPerets2020} for exact, more general expressions.} by a broken power-law
\begin{equation}
  \scri(a_b) \sim \left(\frac{a_b}{a_0}\right)^p.
\end{equation}
The power $p$ depends both on the total angular momentum, and on whether $a_b$ is less than $a_0$ or greater than it.

The critical value of $R_{\max}$ for which the probability of increasing the binary's energy is equal to the probability of decreasing it corresponds to a critical value of $a_{\rm co}$, which satisfies\footnote{Observe that this is not necessarily the same value of $a_{\rm co}$ for which the expected value of $a_b$, $\mathbb{E}(a_b)$, is equal to $a_0$. \emph{This} value may be computed in much the same manner.}
\begin{equation}
  \int_{a_0}^{a_{\rm co}}f(a_b)\mathrm{d}a_b = \int_{0}^{a_0}f(a_b)\mathrm{d}a_b.
\end{equation}
Substituting
\begin{equation}
  p = \begin{cases}
        \frac{1}{2}, & \mbox{if } a < a_0 \\
        \frac{1}{3}, & \mbox{otherwise}.
      \end{cases}
\end{equation}
(a plausible choice for some values of the total angular momentum), yields
\begin{equation}
  a_{\rm co} = a_0\left(\frac{29}{15}\right)^{3/7} \approx 1.326 a_0.
\end{equation}

The relation between $R_{\max}$ and $a_{\rm co}$ is easily derived from energy conservation, where the ejected body is at its apocentre at $R_{\max}$; then,
\begin{equation}
  \frac{GM\mu_s}{R_{\max}} + \frac{Gm_b\mu_b}{2a_{\rm co}} = \frac{Gm_b\mu_b}{2a_0},
\end{equation}
whence
\begin{equation}
  \frac{a_{\rm co}}{a_0} = \frac{1}{1-\frac{4a_0}{R_{\max}}}.
\end{equation}
Solving these equations yields
\begin{equation}\label{eqn:R crit approx}
  \frac{4a_0}{R_{\rm crit}} = 1-\left(\frac{15}{29}\right)^{3/7},
\end{equation}
or $R_{\rm crit} \approx 16.25a_0$.

\begin{figure}
    \centering
    \includegraphics[width=0.45\textwidth]{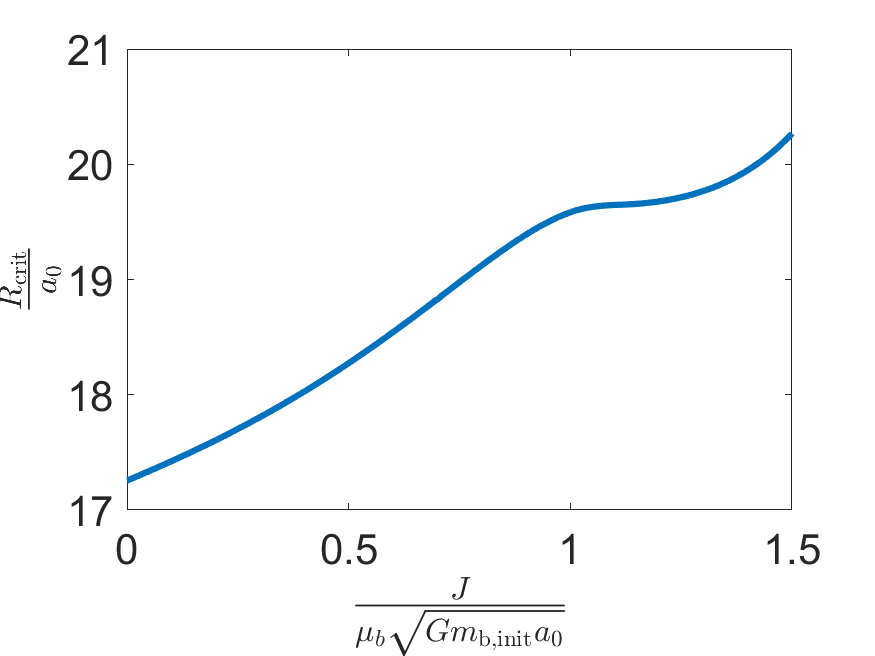}
    \caption{The dependence of the critical value $R_{\rm crit}$ on the total angular momentum of the three-body system. The values are similar to the value of $16.25$ in equation \eqref{eqn:R crit approx}, where now `hard' and `soft' refer to energy changes -- not semi-major axes. The masses and initial conditions are the same as for figure \ref{fig:various R}, except that here the total angular momentum is varied.}
    \label{fig:R_J}
\end{figure}

\begin{figure}
    \centering
    \includegraphics[width=0.45\textwidth]{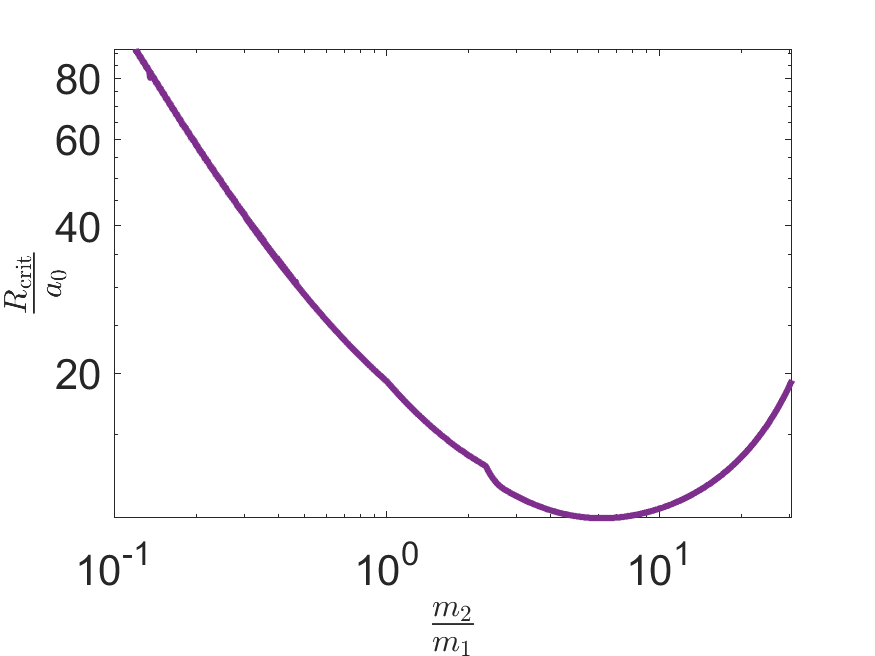}
    \caption{A plot of the critical radius as a function of the mass of one of the initial components of the binary, for the outcome in which the initial components remain in the final binary. Here $R_{\rm crit}$ is defined as the value of $R_{\max}$ (shortened to $R$) for which the probability of increasing the binary energy is equal to that of decreasing it. The masses are $m_1 = m_3 = 1 ~M_\odot$, and the other initial conditions are the same as those of figure \ref{fig:various R}.}
    \label{fig:R_m}
\end{figure}

This value agrees well with an exact (angular-momentum dependent) calculation of $R_{\rm crit}$, using the full formula equation \eqref{eqn:f_bin E_b}, which is shown in figure \ref{fig:R_J}. In figure \ref{fig:R_m}, we show $R_{\rm crit}$ for different mass-ratios.

\begin{figure}
    \centering
    \includegraphics[width=0.45\textwidth]{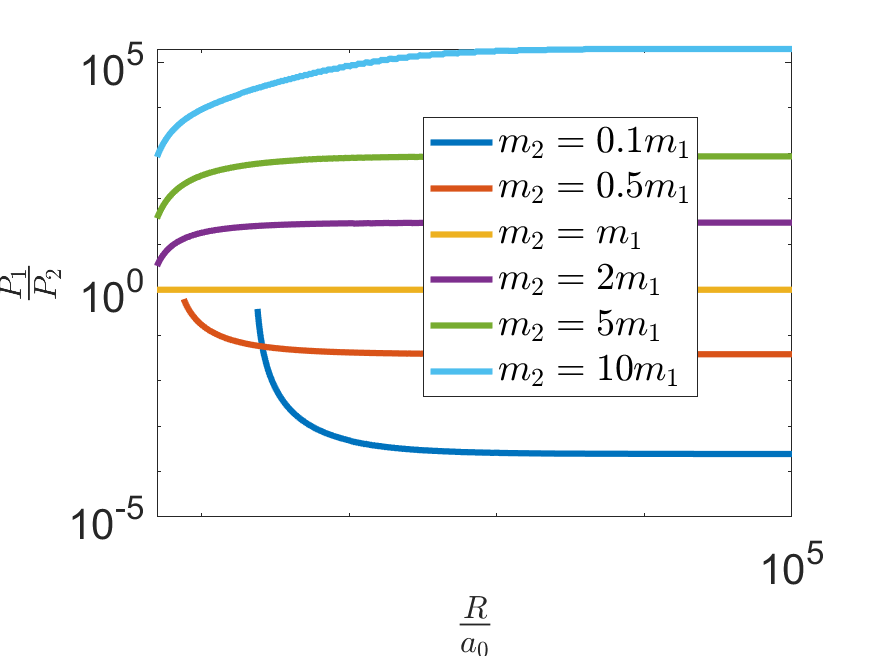}
    \caption{The ratio of the probability of ejecting star $1$ and the probability of ejecting star $2$ in an exchange interaction as a function of $R_{\max}$ (shortened to $R$), for various values of $m_2$ (where $m_1 = m_3 = 1~M_\odot$ and $a_0 = 0.1$ AU and all other parameters are similar to those of figure \ref{fig:various R}). }
    \label{fig:BR_R}
\end{figure}

We also show, in figure \ref{fig:BR_R}, the ratio of the probability of ejecting star $1$ and the probability of ejecting star $2$ in an exchange interaction, for various values of $m_2$. This figure shows that indeed for large values of $R_{\max}$, the probabilities converge to the usual case, where there is no cut-off.

\section{Rate of Hardening}
\begin{figure}
  \centering
  \includegraphics[width=0.45\textwidth]{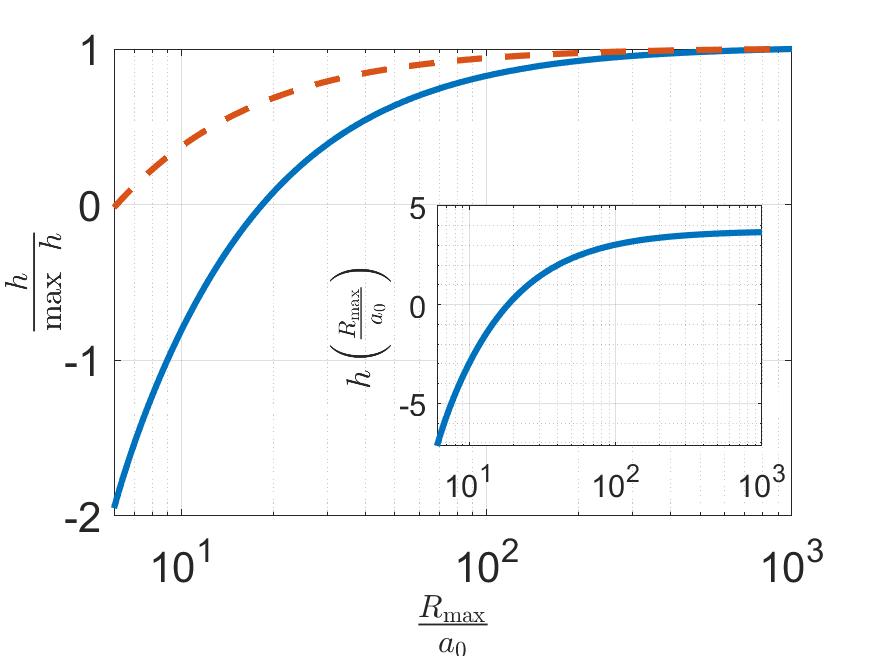}
  \caption{The binary hardening/softening rate, plotted as a function of $R_{\max}$. In the main panel, we plot $h$ normalised by its maximum value, for the equal mass case (blue), and for the case where $m_1 = m_2 = 10~M_\odot$, while $m_3 = 5~M_\odot$ (orange, dashed), conditioned on the event that star $3$ is ejected. In the smaller panel we show $h$ as a function of $R_{\max}$ for the equal mass case, where it can be seen that $\displaystyle \lim_{R_{\max} \to \infty} h = 3.67$. For a given $\frac{R_{\max}}{a_0}$, the value of $\langle \dot{E}_b \rangle$ is independent of $a_0$.}\label{fig:rate}
\end{figure}

We now re-visit the standard calculation of the hardening rate of hard binaries in clusters (\citealt{Binney}, \S 7.5.7c, and \citealt{HeggieHut1993}), and generalise it to include a cut-off; we treat the equal mass case, as it is simple, although the general case is also amenable to the same treatment.
Modifying equation (39) of \cite{HeggieHut1993}, the mean rate of change of the binary's energy is
\begin{equation}
\begin{aligned}
    \langle \dot{E}_b \rangle & = - \frac{4n_sn_b}{3\sqrt{3\pi}\sigma^3}\int_0^\infty \mathrm{d}v_s v_s^3 e^{-v_s^2/3\sigma^2} \int_{D} \frac{2\pi J \mathrm{d}J}{\mu_s^2v_s^2}\\ &
    \times \int \mathrm{d}E_b' (E_b' - E_{b,0})f(E_b'|E_{b,0},J) \\ &
    = - \frac{4n_sn_b}{3\sqrt{3\pi}\sigma^3}\int_0^\infty \mathrm{d}v_s v_s^3 e^{-v_s^2/3\sigma^2} \int_{D} \frac{2\pi J \mathrm{d}J}{\mu_s^2v_s^2}\\ &
    \times \mathbb{E}(\Delta E_b|J,E_{b,0}),
\end{aligned}
\end{equation}
where $\mathbb{E}$ denotes an expectation value, $n_s$ is the cluster's single star number density, $n_b$ -- the binary number density, and $\sigma$ is the cluster's one-dimensional velocity dispersion.\footnote{For simplicty we assume the velocity distribution to be Gaussian. If it isn't, $n_s n_b \exp(-v_s^2/3\sigma^2)$ should be replaced by the appropriate distribution function, which should then be integrated over $v_s$.} The integration over angular momenta, over the allowed range $D$ (divided by $\mu_s^2v_s^2$), turns probabilities into cross-sections and is equivalent to integrating over impact parameters. Let
\begin{equation}
    h \equiv \frac{2\pi}{\mu_s^2 G M a_0\abs{E_{b,0}}} \int_D J\mathrm{d}J ~\mathbb{E}(\Delta E_b|J,E_{b,0}).
\end{equation}
As the total cross-section for hard binary scattering scales like $GMa_0/v_s^2$, it is clear that $h$ is independent of $E_{b,0}$ (recall that, since the binary is hard, the total energy is $\approx E_{b,0}$). In figure \ref{fig:rate}, we plot $h$ as a function of $R_{\max}/a_0$, calculated using equation \eqref{eqn:f_bin E_b}.
Upon performing the integration over the relative velocity, too, we find
\begin{equation}
  \langle \dot{E}_b \rangle = -\sqrt{\frac{3}{\pi}}\frac{G^2 n_sn_b m^3}{\sigma}h\left(\frac{R_{\max}}{a_0}\right).
\end{equation}
One can see that for equal masses, $h \to 3.67$ as $\frac{R_{\max}}{a_0}\to \infty$, yielding a pre-factor of $3.6$, in agreement with the $3.8 \pm 0.2$ predicted by \cite{HeggieHut1993} on the basis of numerical simulations. It emerges from figure \ref{fig:rate}, that this rate drops by about $20\%$ already at $R_{\max}/a_0 = 100$, reaching $0$ at $R_{\max} \approx 18 a_0$. This additional dependence of the hardening rate on the environment is expected to influence theoretical predictions of rates of many astronomical phenomena, such as the binary black-hole merger rate in nuclear stellar discs.

\section{Binaries in Discs}
\label{sec:discs}
For binary planetesimals in planetary systems, such as binary asteroids or binary kuiper-belt objects in the solar system, the tidal force due to the host star (the Sun) provides an external limit, effectively given by the relevant Hill radius, $R_H = r\left(\frac{m_b}{3M_\odot}\right)^{1/3}$, where $r$ is the distance between the initial binary and the Sun, and $M_{\odot}$ is the solar mass. Since one is mainly interested in chaotic encounters, the analysis here pertains chiefly to binary planetesimals in the early solar system (see also \citealt{Perets2011}), or to binary-single encounters of asteroids or trans-Neptunian objects. For example, take a system with equal masses, each equal to half that of the Pluto-Charon system (see, e.g., \citealt{Rozneretal2020,Grishinetal2020} and references therein), which has $m_b \approx 7.3\times 10^{-9}~M_\odot$ and $a_0 = 19571~\textrm{km}$ \citep{Naozetal2010}, and is situated at $r = 39 ~\textrm{AU}$; this system has $R_H = 401 a_0$.
However, in the early solar system, such a binary could have easily existed as close to the Sun as $1 ~\textrm{AU}$, or even much less. That one would have led to $R_H = 10.2 a_0$, which is now below the threshold for softening due to binary-single encounters.
Such a conclusion is only valid if $a_0$ is $\lesssim Gm_b/\sigma^2$, where $\sigma$ is the velocity dispersion.
For a proto-planetary disc, the velocity dispersion grows like $\sigma \propto t^{1/4}$, and starts at $10 ~\textrm{cm s}^{-1}$ \citep{Armitage2010}, which, at the time of formation of planetesimals (at a few million years following the protoplanetary disc formation) ensures that such binaries are hard.

For an Asteroid-Belt binary, like Petit-Prince and Eugenia, whose parameters are $m_b = 2.9\times10^{-12}~M_\odot$, $a_0 = 1180~\textrm{km}$, $r_{\rm peri} = 2.5~\textrm{AU}$ \citep{Naozetal2010}, one finds a mere $R_H = 31a_0$. In the early phases of the solar system, such a binary asteroid would have existed at lower $r$, even $r \approx 1~\textrm{AU}$, where $R_H = 12.5 a_0 < R_{\rm crit}$.

That the effect of a cut-off becomes significant in the collisional evolution of binaries in discs is not accidental. For a spherical system in an external potential, the velocity dispersion and the cut-off $R_{\rm max}$ are generated by the same potential, and so, by the virial theorem, $Gm_b/R_{\max} \sim \sigma^2 \ll Gm_b/a_0$ for a hard binary, whence, for such a binary, $R_{\max} \gg a_0$. The only situation where a binary could be hard, but still with $R_{\max}/a_0$ not very large, is when the velocity dispersion is not generated by the same potential that fixes $R_{\max}$ -- for example, in a disc. Indeed, a disc in a certain background potential is, in general, much cooler than a spherical system with the same size in the same background potential.

Let us mention two more examples of systems where a finite cut-off has a potentially sizeable influence on the hardening rate: nuclear stellar discs around massive black holes (inclding stars in active galactic nuclei, i.e. in AGN discs), and the Galactic disc. For the former, it was already shown previously that the number of encounters needed for a binary to merge due to binary-single encounters depends on the black hole mass \citep{Leighetal2018}; here we can quantify this statement using the updated three-body encounter model. Consider an equal mass binary of mass $m_b = 10~M_\odot$, $0.1$ pc away from a central black hole of mass $M_\bullet = 4 \times 10^6 ~M_\odot$, in a disc with $h/r = 0.01$ \citep{Tagawa_2020}. Then $\sigma = \sqrt{GM_\bullet h^2/r^3} = 4~\textrm{km s}^{-1}$, while $R_H = 194$ AU. This implies that for any $a_0 \gtrsim 9.7$ AU, this binary will soften, in expectation, rather than harden.

For the Galactic thick disc, $h/r \approx 10\%$, while for the thin disc, it is about $3\%$ \citep{Binney}. The cut-off is given by the Hill radius of the Galactic mass interior to the binary's position, i.e. by
\begin{equation}
    R_H = r\left(\frac{m_b}{3M(r)}\right)^{1/3} = r^{4/3}\left(\frac{Gm_b}{v_c^2(r)}\right)^{1/3},
\end{equation}
where $v_c(r)$ is the circular velocity at radius $r$. For $r = r_\odot = 8$ kpc, $v_c(r) \approx 234~\textrm{km s}^{-1}$, whence $\sigma_{\rm thick} = 24~\textrm{km s}^{-1}$, $\sigma_{\rm thin} = 7.3~\textrm{km s}^{-1}$. For such values of the velocity dispersion, it turns out that for plausible values of $a_0$, $m_b$, either $R_H \gg a_0$, or the binary is soft. Therefore, contrary to the above reasoning, whereby the cut-off would be important in discs, it is not so in the Galactic disc. Nevertheless, it might have a significantly  more important effect in bulge-less disc galaxies, where stars can reside in the disc, but be much closer to the centre.

The examples considered in this section are summarised in table \ref{tab:discs}.
\begin{table*}
  \centering
  \caption{\label{tab:discs} A summary of the examples of hard binaries in discs considered in \S \ref{sec:discs}.}
  \begin{tabular}{|c|c|c|c|c|}
  \hline
  Environment & $m_b~ (M_\odot)$ & $r$ & $\sigma ~(\textrm{km s}^{-1})$ & $\frac{R_{\max}}{a_0}$ \\ \hline
  Proto-planetary disc & $7.3 \times 10^{-9}$ & $1$ AU & $3.16\times 10^{-3}$ &  10.2 \\
  Proto-planetary disc & $2.9\times 10^{-12}$ & $1$ AU & $3.16\times 10^{-3}$ & 12.5 \\
  AGN disc & 10 & $0.1$ pc & 4 & 20\\
  Galactic thin disc & $1$ & $8$ kpc & $7.3$ & $>10^4$ \\
  \hline
  \end{tabular}
\end{table*}

\section{Conclusions}

In this paper we showed how to apply the model of \cite{GinatPerets2020} to study the influence of the environment on hard binary-single encounters. We found that this effect can be quite substantial in disc environments, where the cut-off is induced by the external gravitational field, while the velocity dispersion is not (directly) induced by it. In such environments, it is possible for hard binaries not to harden, on average, as in some cases we exemplified. The cut-off changes the binary hardening rate even for large values of $R_{\max}/a_0$, and this change should be accounted for in future theoretical predictions of rates of phenomena involving binary hardening, such as the evolution of binaries in AGN discs, and the rate of binary black-hole mergers in such environments, as well as in studies of the evolution of binary planetesimals during the early evolution of protoplanetary discs.

\section*{Acknowledgements}
Y.B.G. is grateful for the support of the Adams
Fellowship Program of the Israeli Academy of Sciences
and Humanities.

\section*{Data Availability}
The data underlying this article will be shared on reasonable request to the corresponding author.




\bibliographystyle{mnras}
\bibliography{encounters,bibliography}




\bsp	
\label{lastpage}
\end{document}